\documentclass[onecolumn]{svjour2}
\usepackage{graphicx}

\title{The accelerated Kepler problem}

\institute{F. Namouni \at Laboratoire Cassiop\'ee, CNRS UMR 6202\\
              Observatoire de la C\^ote d'Azur, 
              BP 4229, Nice 06304, France\\
              Tel.: +33-492-003023\\
              Fax: +33-492-003118\\
              \email{namouni@obs-nice.fr}           
              \and M. Guzzo \at Dipartimento di Matematica Pura ed Applicata\\
              Universit\`a degli Studi di Padova, 
              via Trieste 63, 35121 Padova,
              Italy\\Tel.: +39-049-8271416\\
              Fax: +33-049-8271204\\\email{guzzo@math.unipd.it}}

\author{F. Namouni \and M. Guzzo}
\date{Received: date / Accepted: date}
\begin{document}

\maketitle

\begin{abstract}
  The accelerated Kepler problem is obtained by adding a constant acceleration
  to the classical two-body Kepler problem. This setting models the dynamics
  of a jet-sustaining accretion disk and its content of forming planets as the
  disk loses linear momentum through the asymmetric jet-counterjet system it
  powers.  The dynamics of the accelerated Kepler problem is analyzed using
  physical as well as parabolic coordinates. The latter naturally separate the
  problem's Hamiltonian into two unidimensional Hamiltonians. In particular,
  we identify the origin of the secular resonance in the accelerated Kepler
  problem and determine analytically the radius of stability boundary of
  initially circular orbits that are of particular interest to the problem of
  radial migration in binary systems as well as to the truncation of accretion
  disks through stellar jet acceleration.  \keywords{Kepler problem \and
    stellar jets \and extrasolar planets \and stellar binaries}
\end{abstract}

\section{Introduction}
The recent studies of the consequences of the acceleration induced by stellar
jets on extrasolar planets and protoplanetary disks (Namouni 2005, 2007,
hereafter Papers I, II; Namouni and Zhou 2006) have shown that the basic
dynamics of these systems rely on the accelerated Kepler problem (hereafter
AKP). Although it is uncommon in celestial mechanics, the problem obtained by
adding a constant acceleration to one of the two bodies in an $r^{-1}$
potential is at the basis of the Stark effect in quantum mechanics (Brandsen
and Joachain 1989). In this case, the acceleration models the constant
electric field applied to an atom.

An interesting property of the AKP is its integrability. This result was
obtained by Epstein (1916) who used the Hamilton-Jacobi appraoch in terms of
parabolic coordinates to separate the AKP into two unidimensional Hamiltonians
(Landau and Lifschitz 1969). The integrability of the AKP in the
position--velocity space was expressed by Redmond (1964) who derived a
generalized Runge-Lenz invariant for the Stark effect. This invariant together
with the energy and the angular momentum component along the direction of
acceleration are the three integrals of the AKP.  Recently, it was recognized
that some of the dynamical consequences of time-variable jet-induced
acceleration on the evolution of accretion disks and the forming planets they
contain can be ascertained from the AKP (i.e. from the dynamics associated
with a time-independent acceleration). To this end, analytical derivations in
terms of the physical variables (positions, velocities, orbital elements) as
well as perturbation theory through secular averaging and sudden impulse
methods were used to determine the eccentricity and inclination excitation of
extrasolar planets (Paper I) as well as the vertical profile of jet-sustaining
accretion disks (Paper II).

The use of parabolic coordinates has the advantage of reducing the AKP into
two integrable unidimensional problems. This however comes at a price:
parabolic coordinates are not physical. Consequently, the aim of this paper is
to link the two analytical approaches of perturbation theory and parabolic
coordinates in order to give a canonical interpretation of the physical
properties described in the works mentioned above.

The paper is organized as follows: in Section 2, we show how the AKP arises in
Keplerian systems with asymmetric momentum loss. In Section 3, we compare some
of the AKP's simple qualitative features to those of the classical Kepler
problem. Section 4 is devoted to the AKP's formulation in parabolic
coordinates. The secular resonance at the basis of the eccentricity excitation
of extrasolar planets is studied in section 5 and its origin in the
non-averaged problem is given in terms of parabolic coordinates.  In section
6, we derive the dependence of the Keplerian boundary radius on the initial
inclination for circular orbits. Section 7 contains further directions of
research in the AKP.

\section{From the classical Kepler problem to the AKP}
Consider a two-body Kepler problem where the larger mass object is loosing
linear momentum at the constant rate $\dot {\bf P}_{\rm loss}$. The equations
of motion of the two bodies in an inertial frame are given as:
\begin{equation}
m\frac{{\rm d}  {\bf v}}{{\rm d} t}=-\frac{GM({\bf
  x-X}) }{|{\bf x-X}|^3}, \ \ \  M\frac{{\rm d} {\bf
    V}}{{\rm d} t}=\frac{Gm({\bf
  x-X})}{|{\bf x-X}|^3} +\dot {\bf P}_{\rm loss},
\end{equation}
where $m,\ M,\ {\bf x,\ X,\ v, \ V}$ are the masses, positions and velocities of the
two bodies and $G$ is the gravitational constant. In terms of the center of
mass and relative motion coordinates, ${\bf x}_r={\bf x-X}$, $(m+M){\bf
  x}_g=m{\bf x}+M{\bf X}$ and the corresponding velocities, ${\bf v}_r,{\bf
  v}_g$, the equations of motion are written as:
\begin{equation}
\frac{{\rm d}  {\bf v}_g}{{\rm d} t}=\frac{\dot {\bf P}_{\rm loss}}{(m+M)}, \
\ \ \frac{{\rm d}  {\bf v}_r}{{\rm d} t}=-\frac{G(m+M)}{|{\bf x}_r|^3}\, {\bf
  x}_r-\frac{\dot {\bf P}_{\rm loss}}{M}.
\end{equation}
These equations show that the motion of the center of mass is accelerated in
the inertial frame by an amount equal to the ratio of the momentum loss rate
to the total mass of the system. The relative motion of the small body remains
independent of the center of mass's motion while it is accelerated in the
opposite direction to momentum loss by the ratio of the momentum loss rate to
the central mass.

Calling ${\bf A}=- \dot{\bf P}_{\rm loss}/M$ the constant acceleration of the
relative motion and dropping the index $r$ for simplicity, the relative motion
of the smaller object around the larger one is given by:
\begin{equation}
\frac{{\rm d} {\bf v}}{{\rm d} t}=-\frac{k}{|{\bf x}|^3}\,{\bf
  x} +{\bf A}, \label{motion}
\end{equation}
where $k=G(M+m)$. This equation defines the accelerated Kepler problem. It
describes the dynamical evolution of a planet embedded in a jet-sustaining
disk that looses mass and linear momentum from its inner parts (the so-called
jet launching region, Paper II). 

\section{Constants of motion and equilibrium points}

As the perturbing acceleration is constant, the dynamical problem described by
(\ref{motion}) remains conservative and amounts to the addition of the
potential $R={\bf A}\cdot {\bf x}$ to the two-body Kepler problem. This in
turn gives the energy integral of the AKP as:
\begin{equation}
E=\frac{v^2}{2}-\frac{k}{|{\bf x}|} -{\bf A}\cdot {\bf x}. \label{energy}
\end{equation}
The acceleration ${\bf A}$ defines a preferred direction in the AKP. As a
result, the projection of the specific angular momentum ${\bf h}={\bf
  x}\times{\bf v}$ along this direction,
\begin{equation}
h_z={\bf h}\cdot {\bf A}/A,\label{hz}
\end{equation} is
conserved as $\dot {\bf h}={\bf x}\times {\bf A}$. The third integral (Redmond
1964) is given as:  
\begin{equation}
\beta= k \, \frac{{\bf A}\cdot {\bf  e}}{A} +\frac{({\bf x}\times {\bf A})^2}{2\,A}, \label{beta}
\end{equation}
where ${\bf e}={\bf v}\times {\bf h}/k-{\bf x}/|{\bf x}|$ is the
Runge-Lenz vector. The constant $\beta$ replaces the Runge-Lenz vector of the
classical Kepler problem.

To further compare the AKP with the classical Kepler problem, we examine its
orbits of least energy. As the acceleration imposes a preferred direction, we
choose a reference frame where the $z$--axis is directed along ${\bf A}$.
Using the conservation of $h_z$, the energy integral (\ref{energy}) can be
written as:

\begin{equation}
E=\frac{\dot \rho^2+\dot z^2}{2}+W(\rho,z), \ \ \mbox{where} \ W(\rho,z)=\frac{h_z^2}{2\rho^2}-\frac{k}{\sqrt{\rho^2+z^2}}
-Az, \label{energy2}
\end{equation}
and where $\rho$ and $z$ are the usual cylindrical coordinates in the plane
orthogonal to ${\bf A}$.  For a given value of $h_z$, the zero-velocity curves
(level curves of $W$) are shown in Fig. 1 for the classical Kepler problem and
in Fig. 2 for the AKP. For the classical Kepler problem, only a stable
equilibrium associated with circular motion is present. In the case of the
AKP, the addition of the acceleration potential and its odd-symmetry introduce
a new unstable equilibrium as well as a motion separatrix. The two equilibrium
orbits (stable and unstable) of the AKP are located above the reference plane,
$z=0$, that contains the central object.  The separatrix denotes the boundary
where the acceleration from the central object matches the amplitude of the
perturbing acceleration. Motion outside the separatrix is unbounded.
\begin{figure}
\begin{center}
\includegraphics[width=0.6\textwidth]{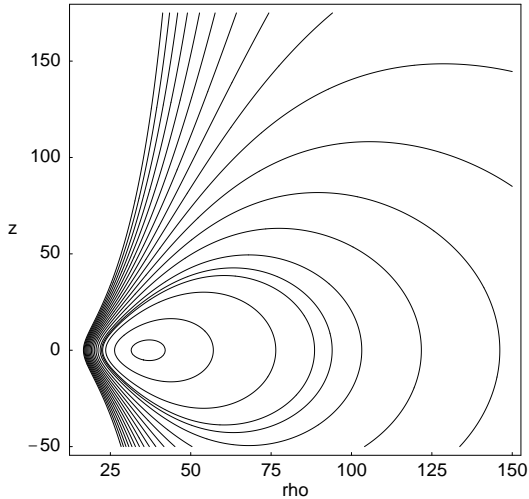}
\caption{Zero-velocity curves of the classical Kepler problem in the
  cylindrical coordinate plane ($\rho,z$). The unit distance is the
  Astronomical Unit (AU). The central object's mass is equal to the Sun's and
  the value of $h_z$ used in the expression of $W$ is that of a planar
  circular orbit at $\rho=36\,$AU. }
\end{center}
\label{fig-1}
\end{figure}
\begin{figure}
\begin{center}
\includegraphics[width=0.6\textwidth]{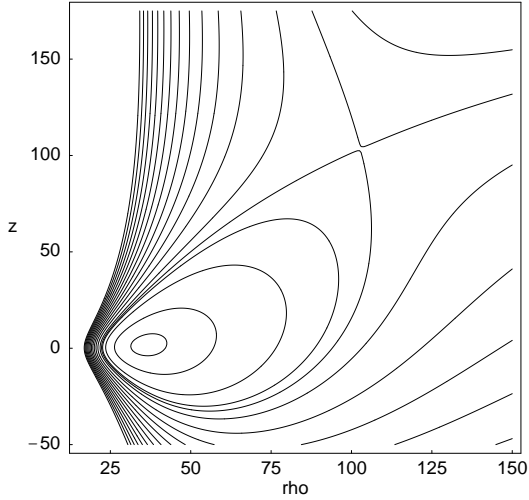}
\caption{Zero-velocity curves of the AKP in the
  cylindrical coordinate plane ($\rho,z$). The unit distance is the
  Astronomical Unit (AU). A constant vertical acceleration of amplitude
  $A=2\times 10^{-10}$\,km\,s$^{-2}$ was added to the classical Kepler problem
  of Fig. 1. The odd-symmetry of the acceleration's potential
  removes the reflection symmetry of the Kepler problem with respect to
  $z$. The region of bounded motion is restricted by the appearance of the
  unstable orbit at $(\rho,z)=(102.8,103.5)$.}
\end{center}
\label{fig-2}
\end{figure}
To determine the location of the equilibrium points, we minimize the energy
integral with respect to $\dot\rho, \ \dot z, \ \rho,$ and $ z$. The first two
derivatives yield $\dot \rho=\dot z=0$ showing that least energy orbits are
circular and planar. The derivative with respect to $\rho$ yields the orbital
rotation rate as a function of the projection of the angular momentum on the
direction of acceleration, $h_z$. Using the fact that orbits are planar and
circular,  $h_z=n \rho^2$ and the orbital rotation rate:
\begin{equation}
n=
\left[\frac{k}{\left(\rho^2+z^2\right)^{3/2}}\right]^\frac{1}{2}.
\label{sombrero1}
\end{equation}
The derivative with respect to $z$ amounts to equating the vertical
acceleration, ${\bf A}$, and the vertical pull of the central object and
yields
\begin{equation}
\rho=\left[\left(\frac{k\, z}{A}\right)^\frac{2}{3}-z^2\right]^\frac{1}{2}.
\label{sombrero}
\end{equation}
The circular orbits hover above the central object increasingly higher with
distance. In other words, a circular orbit in the AKP is still planar but the
corresponding orbital plane no longer contains the central object. The latter
becomes increasingly distant from that plane as the orbital radius increases.
Figure (3) shows the family of circular orbits of the AKP in the $\rho
z$--plane.  The uppermost orbit is located at $\rho=0$ and $z=\sqrt{k/A}$.

The locus of least energy orbits given by equations (\ref{sombrero1}) and
(\ref{sombrero}) does not indicate where the stable orbits end and the
unstable orbits start.  To find this limit, we express the specific angular
momentum, $h_z$, for the least energy orbits (\ref{sombrero}) as a function of
$z$. As the equilibria of $W$ in the $\rho z$--plane are determined by the
same $h_z$, their coordinates are found by solving for $z$ at constant $h_z$
and using equation (\ref{sombrero}) to determine $\rho$. The maximum of
$h_z(z)$ is reached at:
\begin{equation}
z_{\rm crit}= \left[\frac{k}{27 A}\right]^\frac{1}{2} \ \ \mbox{and} \ \ \
\rho_{\rm crit}= \left[\frac{8 k}{ 27 A}\right]^\frac{1}{2}. \label{zcrit}
\end{equation}
This location determines the last stable planar circular orbit of the AKP.
Figure (4) shows the function $h_z^2(z)$ as well as $\rho(z)$
(\ref{sombrero}). The maximum of $h_z$ is where the stable and unstable points
meet; above this value of $h_z$ motion is unbounded in all space.  For $z<
z_{\rm crit}$, planar circular orbits are stable and their profile in the
$\rho z$--plane has the shape of a sombrero curved along the direction of
acceleration (Fig.  3).  In Paper II, the sombrero profile was identified as
the instantaneous state of least energy of a jet-sustaining disk. This profile
was used to ascertain the changes in the temperature, hydrostatic equilibrium
and heating of the disk.

\begin{figure}
\begin{center}
\includegraphics[width=0.65\textwidth]{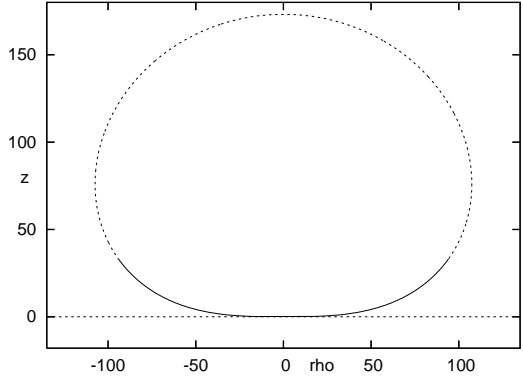}
\caption{Least energy orbits of the AKP in the $\rho z$--plane. The system's parameters are those of
  Fig. 2. The solid line stands for the stable equilibrium orbits (the sombrero
  profile) and the dashed line for the unstable equilibrium orbits (see Fig. 4
  for more details). }
\end{center}
\label{fig-3}
\end{figure}

\begin{figure}
\begin{center}
\includegraphics[width=0.65\textwidth]{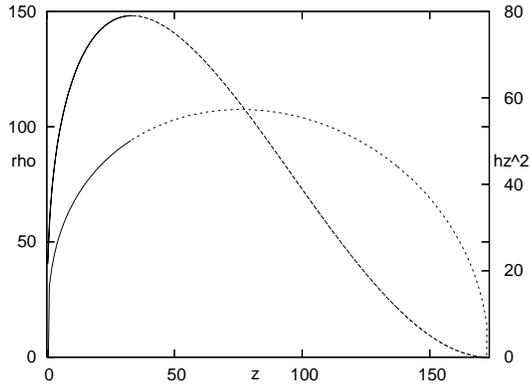}
\caption{The identification of the equilibrium orbits' stability through the
  vertical component of angular momentum. The system's parameters are those of
  Fig. 3. Least energy orbits (lower maximum curve) are shown in the
  $z\rho$--plane (left scale). The function $h_z^2(z)$ (higher maximum curve,
  right scale) is determined from the equilibrium orbits (Eqns.
  \ref{sombrero1} and \ref{sombrero}) and normalized to the angular momentum
  of a circular orbit at 1\,AU.  The function $h_z^2(z)$ yields two values of
  $z$ for a given value of $h_z$. These are the stable (solid line) and
  unstable (dashed line) equilibria of Fig. 2.  The maximum of $h_z^2(z)$
  denotes the outer edge of the sombrero profile.}
\end{center}
\label{fig-4}
\end{figure}

Lastly, we remark that the existence of the constant of motion $\beta$ shows
that the separatrix of $W$ is pendulum-like and does not involve any kind of
chaotic motion. Orbits that are set up near the unstable point such as that in
Figures 7 and 8 of Paper I tend to escape after a long time (relative to the
orbital period) only because of the discrete nature of the numerical scheme
used to integrate the orbit. In section 6, we discuss further the outer
boundary of initially circular orbits whose orbital plane is inclined to the
direction of acceleration such as those studied in Paper I. Before doing so,
we first introduce the formulation of the AKP in parabolic coordinates that
will be needed in sections 5 and 6.

\section{The AKP in parabolic coordinates} 
Parabolic coordinates ($\xi,\eta,\theta$) are related to the cylindrical
coordinates ($\rho,z,\theta$) through (Landau and Lifschitz 1969):
\begin{equation}
z=\frac{1}{2}\,(\xi-\eta), \ \ \ \rho=\sqrt{\xi\,\eta} \label{paracor}.
\end{equation}
In terms of the orbital radius $r^2=\rho^2+z^2$, these coordinates are written
as:
\begin{equation}
\xi=r+z, \ \ \ \eta=r-z \label{paracor2}. 
\end{equation}
The conjugate momenta of  parabolic coordinates are given by:
\begin{equation}
p_\xi=\frac{\dot\xi}{4\,\xi}\ (\xi+\eta), \ \ \ p_\eta=\frac{\dot\eta}{4\,\eta}\ (\xi+\eta), \ \ \ p_\theta=\xi\,\eta\,\dot\theta.
\end{equation}
The Hamiltonian of the AKP can now be written in terms of the parabolic
coordinates and momenta as:
\begin{equation}
H=2\,\frac{\xi\, p_\xi^2+\eta\,
  p_\eta^2}{\xi+\eta}+\frac{p_\theta^2}{2\,\xi\,\eta}-\frac{2k}{\xi+\eta} -\frac{A}{2}\,
  (\xi-\eta),\label{hamil}
\end{equation}
where we used the relations (\ref{paracor}) and (\ref{paracor2}) in the
expressions of the gravitational potential as well as the acceleration
potential $R=A\,z$ (third and fourth terms on the right hand side). As the
Hamiltonian is autonomous and independent of $\theta$, we recover the two
constants of motion: the energy (\ref{energy}), $H=E$, and the angular
momentum component along the direction of acceleration (\ref{hz}),
$p_\theta=h_z$. By applying the Hamilton-Jacobi approach with the separation
of variables method, it can be shown (Epstein 1916, Landau and Lifschitz 1969)
that the Hamiltonian (\ref{hamil}) decouples into two unidimensional
Hamiltonians that are related by the energy $E$, the vertical component of
angular momentum, $h_z$ and a third integral, $\beta$ defined in equation
(\ref{beta}), and given by:
\begin{equation}
\beta=2\,\eta\, p_\eta^2+\frac{h_z^2}{2\,\eta}+\frac{A\,\eta^2}{2}-E\,\eta-k,\ \ \ \ 
-\beta=2\,\xi\, p_\xi^2+\frac{h_z^2}{2\,\xi}-\frac{A\,\xi^2}{2}-E\,\xi-k.  \label{beta2}
\end{equation}
The two unidimensional Hamiltonians are therefore:
\begin{eqnarray}
E&=&2\,p_\eta^2+\frac{h_z^2}{2\,\eta^2}+\frac{A\,\eta}{2}-\frac{\beta+k}{\eta},\label{hm1}\\ 
E&=&2\,p_\xi^2+\frac{h_z^2}{2\,\xi^2}-\frac{A\,\xi}{2}+\frac{\beta-k}{\xi}. \label{hm2}
\end{eqnarray}
The choice of vertical direction for the coordinate frame along the direction
of acceleration implicitly imply that $A$ is positive. Close examination of
(\ref{hm1}) and (\ref{hm2}) shows that it is (\ref{hm1}) that has a potential
function (last three terms on the right hand side) that has a minimum
regardless of any constants because of the positive linear term associated
with $A$. When motion is possible, orbits are bounded. Motion is possible only
if $E$ is larger than the potential function minimum corresponding to a value
that we shall call $\eta_{\rm s}$. The motion in the $\eta p_\eta$--plane is
shown in Figure (5).

\begin{figure}
\begin{center}
\includegraphics[width=0.75\textwidth]{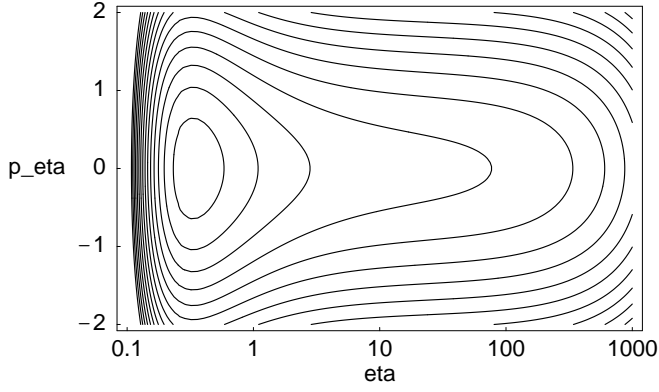}
\caption{Contour plots of the Hamiltonian (\ref{hm1}) in the
  $\eta\,p_\eta$--plane with the parameters $h_z=1$, $A=0.1$, $\beta=1$ and
  $k=2$. Motion is always bounded.}
\end{center}
\label{fig-5}
\end{figure}

Motion in the $\xi p_\xi$--plane depends on the constants of motion as the
negative linear acceleration term needs to be compared to the other potential
terms that appear in the potential function:
\begin{equation}
W_\xi=\frac{h_z^2}{2\,\xi^2}-\frac{A\,\xi}{2}+\frac{\beta-k}{\xi}\label{xi-pot}.
\end{equation}
The derivative of $W_\xi$ shows that the extrema satisfy the polynomial
equation:
\begin{equation}
A\,\xi^3+2\,(\beta-k)\,\xi+2\,h_z^2=0. \label{xi-pot-min}
\end{equation}
For $\beta\geq k$, this equation has no solution (as $A$ is positive) and
$W_\xi$ is a monotonic decreasing function. Consequently, motion is unbounded.
For $\beta< k$, there are extremum points only if:
\begin{equation}
h_z^2<\left[\frac{3}{2}\, (k-\beta)\right]^\frac{3}{2} A^{-\frac{1}{2}}.
\end{equation}
In this case $W$ has a relative (stable) minimum and a relative (unstable)
maximum that we call $\xi_{\rm s}$, and $\xi_{\rm u}$ respectively.  Bounded
motion occurs for $W_\xi(\xi_{\rm s})\leq E<W_\xi(\xi_{\rm u})$; otherwise
orbits are unbounded. Motion in the $\xi\, p_\xi$--plane is shown in Figure
(6). The level curves of $W_\xi$ bear a strong resemblance to the
zero-velocity curves of the AKP in cylindrical coordinates (Fig. 2). Whereas
the latter curves represent motion boundaries for a given $h_z$ and $E$, the
former correspond to actual trajectories.  Lastly, we note that the
equilibrium orbits derived in section 3, correspond to $(\xi_{\rm s},\eta_{\rm
  s})$ (stable, sombrero profile) and $(\xi_{\rm u},\eta_{\rm s})$ (unstable).

\begin{figure}
\begin{center}
  \includegraphics[width=0.75\textwidth]{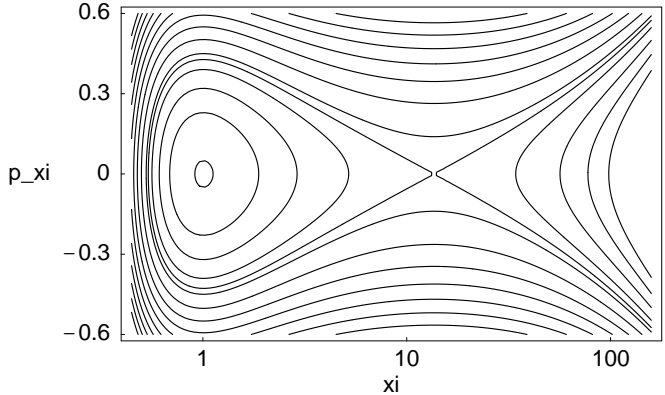}
\caption{Contour plots of the Hamiltonian (\ref{hm2}) in the
  $\xi\,p_\xi$--plane with the parameters  $h_z=1$, $A=0.1$, $\beta=1$ and
  $k=2$. The separatrix passing through the unstable point at $\xi_{\rm u}$
  divides space into bounded and unbounded motion.}
\end{center}
\label{fig-6}
\end{figure}

\section{Secular resonance}

The physical setting of the AKP is useful in identifying the dynamical
response of a planet embedded in a accretion disk that sustains an asymmetric
jet system. Such planets form inside the gas disk on circular planar orbits.
As the disk's state of least energy is the sombrero profile (Paper II), the
formed planet may assume one of the circular sombrero orbits and is therefore
immune to any additional dynamical excitation from the acceleration of the
star-disk system. In order to account for the large eccentricity of extrasolar
planets through the AKP, the ``initial'' planetary orbital plane must be
inclined with respect the direction of acceleration in order to display the
librations described by the two Hamiltonians (Eqn. \ref{hm1} and \ref{hm2}).
The possible physical realization of this geometry is when the jet system is
inclined with respect to the star's rotation axis and therefore precesses
around it. In this case, the acceleration has a constant vertical component
that defines the sombrero profile and a precessing component that may excite
the eccentricity of a planet embedded in the disk.  This was the premise of
Paper I where the excitation of eccentricity and inclination of extrasolar
planets that formed in a jet-sustaining disk was examined.

To illustrate the eccentricity excitation of a planet on an initially circular
planar orbit, it is instructive to first consider the situation where the
precession frequency is so small that the acceleration appears to maintain a
constant inclination with respect to the planetary orbit. In the context of the
AKP, studying the excitation due to an inclined constant direction
acceleration amounts to choosing initially inclined circular planetary orbits
in the presence of a  vertical acceleration.

To examine this situation, the AKP was viewed in Paper I as a classical
two-body Kepler problem perturbed by the acceleration ${\bf A}$.  Working in
the secular regime where the excitation time $|{\bf v}|/A$ is much larger than
the orbital period $2\pi/n$, the acceleration potential $R={\bf A}\cdot {\bf
  x}$ was averaged over the orbital period to find:
\begin{equation}
\left<R\right> =-\frac{3}{2}
 a \, {\bf A}\cdot {\bf e}=-
\frac{3}{2}
A \, a e \sin(\varpi-\Omega) \sin I
\label{secpot}
\end{equation}
where $e$ and $a$ are the orbital eccentricity and semi-major axis.  The
last expression is obtained by choosing the $z$--axis along ${\bf A}$; in this
case, $\varpi$, $\Omega$ and $I$ are the longitude of pericenter, the
longitude of ascending node and the inclination of the angular momentum vector
${\bf h}$ with respect to the acceleration ${\bf A}$. The disappearance of the
longitude in the perturbing potential implies that the semi-major axis is
constant throughout the evolution.

In terms of orbital elements, the conservation of $h_z$ can be written as
$\sqrt{1-e^2} \cos I = \cos I_0$, where $I_0$ is the initial inclination of
the circular orbit. This allows us to reduce the problem to a single degree of
freedom with the equations:
\begin{equation}
\dot e=-\frac{\sqrt{1-e^2}}{na^2\,e}\ \frac{\partial\left< R\right>}{\partial
     \omega}, \ \ \ \dot \omega=\frac{\sqrt{1-e^2}}{na^2\,e}\
     \frac{\partial\left< R\right>}{\partial e}, \label{edotomdot}
\end{equation} 
where $\omega=\varpi-\Omega$ is the argument of pericenter and
$n=\sqrt{k/a^3}$ is the mean motion. The secular potential is now given as:
\begin{equation} \left< R\right>= -\frac{3}{2} A\, a \,
  \sqrt{\frac{\sin^2I_0-e^2}{1-e^2}}\, e \sin \omega. \label{secpot2}
\end{equation}
The orbital elements $e$ and $\omega$ therefore follow curves of constant
$\left< R\right>$. In particular, the potential $\left< R\right>$
(\ref{secpot2}) has extremum points at:
\begin{equation}
\omega=\pm 90^\circ,\ \ 
e=\sqrt{2} \sin (I_0/2), \ \ \mbox{and}\ \ I=\cos^{-1}(\sqrt{\cos I_0}).
\label{exactres}
\end{equation}
This equilibrium may be interpreted as a secular resonance where the
precession rates of the pericenter and node become equal. The maximum
value of $e$ is $\sin I_0$ and corresponds to the cycle of initially circular
orbits. The curves of constant $\left< R\right>$ are shown in Figure (7) for
an initial inclination of $I_0=45^\circ.$

\begin{figure}
\begin{center}
\includegraphics[width=0.75\textwidth]{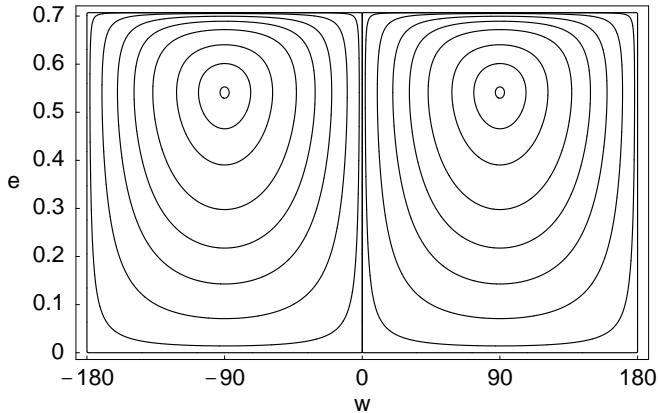}
\caption{Contour plots of the averaged acceleration potential $\left<
    R\right>$ (\ref{secpot2}) for an initial inclination $I_0=45^\circ$. }
\end{center}
\label{fig-7}
\end{figure}
  
The value of $\left< R\right>$ being constant throughout the evolution, it is
possible to derive the analytical expression of the eccentricity evolution.
Denoting by $C$ the value of $-2\left< R\right>/3Aa$, the equations
(\ref{edotomdot}) yield after some algebra:
\begin{equation}
e^2(t)=\left[\left(\frac{\sin^2 I_0+C^2}{2}\right)^2-C^2\right]^\frac{1}{2}\,
\cos\left[ \frac{3A}{na}\ t +\phi\right]+\frac{\sin^2 I_0+C^2}{2}, 
\label{sol-e}
\end{equation}
where $\phi$ is a phase corresponding to the initial value of $e$.  The
expression of $\sin\omega$ as a function of time can be deduced from the
secular potential (\ref{secpot2}). For initially circular orbits, $C=0$ and
$\omega=0$ modulo $180^\circ$.  In this case, equation (\ref{sol-e}) reduces
to the expression given by Equation (8) of Paper I:
\begin{equation}
e(t)= \left|\sin\left[ \frac{3A}{2na}\ t\right]\ \sin I_0 \right|.
\label{eoft}
\end{equation}

Equation (\ref{sol-e}) shows that the eccentricity time evolution occurs at
the excitation frequency:
\begin{equation}
n_A=\frac{3A}{na},
\end{equation}
and that all libration cycles  around exact secular resonance possess this same
frequency. This was shown to be the case in Paper I where orbits were
integrated numerically with the full equations of motion (i.e. without
averaging) and a good match was found between the numerical evolution and
equation (\ref{eoft}) for initially circular orbits. 

To understand the origin of the secular resonance in terms of parabolic
coordinates, we numerically integrated the equations of motion (\ref{motion})
for orbits that were initially placed at exact resonance as defined by the
osculating orbital elements given by (\ref{exactres}). The result is shown in
Figure (8). It turns out that exact resonance orbits are on the stable
equilibrium $\xi_{\rm s}$ in the $\xi p_\xi$--plane whereas the corresponding
$\eta p_\eta$--motion has a finite libration amplitude. This correspondence
allows us to derive a geometric property of secular resonance orbits. As
$\dot\xi=0$, $r+z$ (Eqn. \ref{paracor2}) is constant for resonant orbits.
This in turn shows that resonant motion is confined to the surface of the
parabolic paraboloid:
\begin{equation}
z=\frac{\xi_0^2-\rho^2}{2\xi_0},
\end{equation}
where $\xi_0$ is the initial value of $\xi$.

\begin{figure}
\begin{center}
\includegraphics[width=0.65\textwidth]{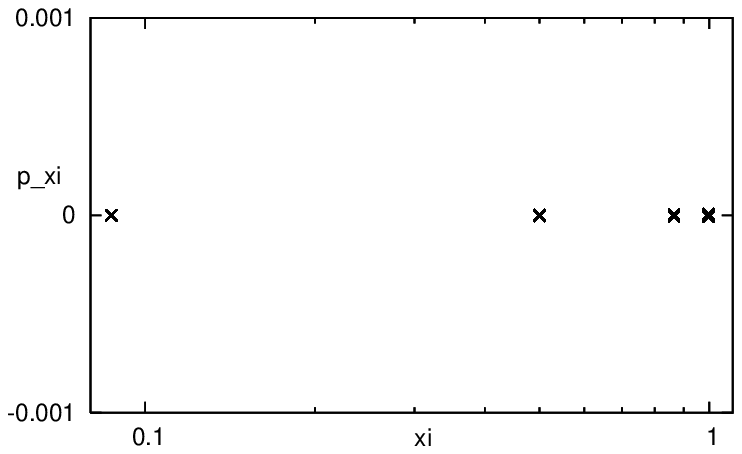}\\
~~~~\includegraphics[width=0.613\textwidth]{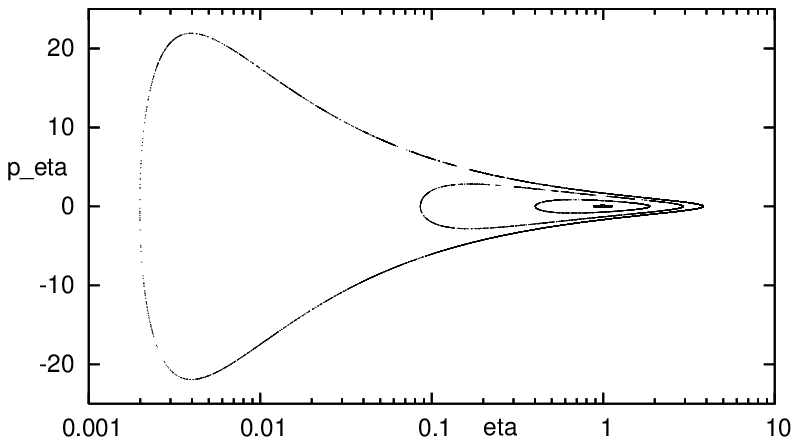}
\caption{Resonance orbits in parabolic coordinate phase space. The AKP
  parameters are those of a test particle orbiting a solar mass star
  accelerated by $A=5\times 10^{-11}$\, km\,s$^{-2}$. All orbits share the
  same initial osculating semi-major axis $a=1$\,AU, argument of pericenter
  $\omega=90^\circ$, longitude of ascending node $\Omega=0$ and true longitude
  $\theta=0$. The different orbits correspond to different initial
  inclinations: $I_0=5^\circ,\ 30^\circ,\ 60^\circ$ and $85^\circ$; their
  eccentricities are given by the expressions (\ref{exactres}).  The
  equilibrium position along $\xi$-axis is larger for smaller $I_0$ while the
  larger $I_0$, the larger the $p_\eta\eta$--amplitude. The units are AU for
  $\xi$ and $\eta$, and AU\,year$^{-1}$ for $p_\xi$ and $p_\eta$.}
\end{center}
\label{fig-8}
\end{figure}

\section{Keplerian boundary of initially circular orbits}
The notion of a boundary in the AKP results from the possible equality at some
relative distance of the two bodies of the gravitational interaction and the
perturbing acceleration. Unlike the scale invariant classical two-body Kepler
problem, the dimension of $A$ naturally gives rise to a typical length that is
proportional to $(k/A)^{1/2}$. In Paper I, it was recognized that this length
scale had to be an upper bound of the region of bounded motion and that a more
precise estimate of the boundary is given by the equality of the mean motion
$n$ and the excitation frequency $n_A$. This equality yields the limit beyond
which the averaging procedure used in the derivation of $\langle R\rangle$
(Eqn. \ref{secpot}) breaks down. The corresponding radius is given as:
\begin{equation}
a_{\rm kplr}=\left[\frac{k}{3A}\right]^\frac{1}{2}\simeq 10^3\,
\left[\frac{2\times 10^{-12}\ {\rm km}\, {\rm s}^{-2}}{A}\right]^\frac{1}{2}\, \left[\frac{M}{M_\odot}\right]^\frac{1}{2}\, {\rm AU}.\label{akplr}
\end{equation}
In terms of $a_{\rm kplr}$, the orbit that defines the outer edge of the
sombrero profile (\ref{zcrit}) is given as:
\begin{equation}
z_{\rm crit}= \frac{a_{\rm kplr}}{3}\ \ \mbox{and} \ \ \
\rho_{\rm crit}=\frac{2\sqrt{2}}{3} \ a_{\rm kplr}.
\end{equation}
Using the formulation of the AKP in parabolic coordinates, it is possible to
solve for the radius of the outer boundary of initially circular orbits whose
plane normal is inclined to the direction of acceleration such as the orbits
of the previous section and those considered in Paper I. These orbits are
circular and planar but do not belong to the sombrero profile (i.e. are not
least energy orbits) as they are initially inclined. For bounded motion to be
possible, the Hamiltonian (\ref{hm2}) must have two equilibria. This in turn
implies that (\ref{hm2}) has three roots at $p_\xi=0$ for a given initially
circular but inclined orbit. These roots are the solution of the equation:
\begin{equation}
E=\frac{h_z^2}{2\,\xi^2}-\frac{A\,\xi}{2}+\frac{\beta-k}{\xi}. \label{hm21}
\end{equation}
Calling $a$ and $I_0$ the initial osculating semi-major axis and inclination
of the circular orbit and taking $k=1$, the constants that appear in
(\ref{hm21}) are given as:
\begin{equation}
E=-\frac{1}{2a},\ h_z^2=a \cos I_0, \ A=\frac{1}{3a_{\rm kplr}^2},
\ \mbox{and}\  \beta=\frac{1}{6}\left(\frac{a}{a_{\rm kplr}}\right)^2.  
\end{equation}
By substituting these constants into (\ref{hm21}) and writing $\alpha =
(a/a_{\rm kplr})^2/6$ and $\xi= \Xi a_{\rm kplr}$, the three roots may be
obtained from the cubic equation:
\begin{equation}
\alpha \Xi^3-\frac{\Xi^2}{2}+(1-\alpha) \Xi - \frac{\cos^2I_0}{2}=0.  \label{hm22}
\end{equation}
The radius of the stable boundary is derived from (\ref{hm22}) as the value of
$\alpha$ for which the discriminant of the cubic equation vanishes; in this
case, the equation does not admit three distinct real roots. This condition
reads:
\begin{equation}
\alpha^4-3 \alpha^3+\left(49 - 18
  \cos^2I_0-27\cos^4I_0\right)\,\frac{\alpha^2}{16}-\frac{9 \alpha
  \sin^2I_0}{8}+\frac{\sin^2I_0}{16}=0. \label{quartic}
\end{equation}
The solution of the quartic equation can be obtained by the usual standard
methods; however, the expressions of the roots are quite cumbersome and are
not given here explicitly. Instead, we use the solution of Eqn.
(\ref{quartic}) directly to plot in Figure (9) the size of the stable
boundary, $a$, normalized to $a_{\rm kplr}$, (or equivalently $\alpha$) as a
function of the initial inclination $I_0$. The region of bounded motion occurs
inside $0.72\, a_{\rm kplr}$ for all inclinations and its boundary decreases
with increasing inclination, $I_0$. 

In Paper I, two examples of escape orbits were given: first, for $a_{\rm
  kplr}=100$\,AU an orbit was initially placed at $a_0=68.5$\,AU with an
inclination of $I_0=30^\circ$ (Figures 7 and 8 of Paper I). At such an
inclination, Figure (9) yields a boundary radius of $a_0=68.5$\,AU. The chosen
orbit was therefore on the unstable equilibrium point of the
$\xi$--Hamiltonian. The second example was that of the stellar binary
companion to $\upsilon$ Andromedae. To illustrate how to reach the current
orbital configuration of this multiplanet binary system, a time dependent
acceleration with a minimum $a_{\rm kplr}=300$\,AU was applied to the
configuration where the stellar companion was initially placed at $a=298$\,AU
with $I_0=60^\circ$. In this case, Figure (9) shows that the minimum stable
boundary radius corresponds to $a=198$\,AU. As the acceleration pulse was
reaching its maximum, the stable boundary radius decreased from infinity and
approached 198\,AU.  The stellar companion was therefore no longer
gravitationally bound to $\upsilon$ Andromedae and moved away from it (Figure
14 of Paper I). However as the orbital time of the companion's motion, $\sim
5000$\,years, was larger than the acceleration decay timescale of
$2000$\,years, the stellar companion was not left with enough time to escape
before the boundary radius bounced back to infinity. The net result of the
acceleration led to the radial migration as well as the eccentricity and
inclination excitation of the stellar companion.

\begin{figure}
\begin{center}
\includegraphics[width=0.65\textwidth]{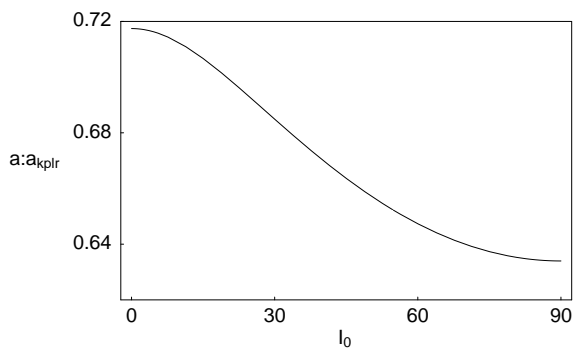}
\caption{The size of the Keplerian boundary of initially circular and inclined
  orbits in the AKP. The ratio of $a$ to $a_{\rm kplr}$ is shown as a function
  of $I_0$ (degrees). }
\end{center}
\label{fig-9}
\end{figure}

\section{Further developments}
In this paper, we have examined the dynamics of the accelerated Kepler problem
in detail using the problem's integrability. We have (i) compared the AKP to
the classical Kepler problem, (ii) derived analytically the features of least
energy orbits, (iii) derived the general solution of the secular problem and
interpreted the secular resonance in terms of parabolic variables. Lastly
(iv), we used the formulation of parabolic variables to determine analytically
the boundary of initially inclined circular orbits that are of particular
importance to the problem of radial migration in binary systems as well as to
the truncation of accretion disks through stellar jet acceleration.

The AKP's integrability makes it an interesting Hamiltonian system that can be
perturbed by additional processes. The perturbation may result in the loss of
integrability and may lead to possible diffusion. In turn, such mechanisms may
further enhance eccentricity excitation and radial migration.  In the context
of the dynamics of forming protoplanets in a jet-sustaining disk, the modeling
of a general time dependence of the perturbing acceleration provides a more
realistic model of jet-disk interaction with further consequences such as disk
heating (simples examples are given in Papers I and II).

An interesting simple dynamical problem that currently remains unsolved is the
possible integrability or quasi-integrability of the precessing AKP. The
resonance studied in section 5 illustrated how to excite the eccentricity of
an initially circular orbit in the case where the constant-amplitude
acceleration has a small precession frequency compared to the excitation
frequency $n_A$. The more general case of a precessing constant-amplitude
acceleration (that corresponds to a precessing jet) yields the precessing AKP
equations:
\begin{equation}
\frac{{\rm d} {\bf v}}{{\rm d} t}
=-\frac{k}{|{\bf x}|^3}\,{\bf
  x} +A\,{\bf u}, \ \ \mbox{and} \ \ \frac{{\rm d} {\bf u}}{{\rm d} t}={\bf \Omega}_A\times {\bf u}\label{pakp},
\end{equation}
where ${\bf \Omega}_A$ denotes the acceleration's constant rotation vector,
$A$ is constant and ${\bf u}$ is the precessing direction of acceleration. To
recover the setting of the AKP, two rotations may be applied to (\ref{pakp}).
First, a rotation at the rate $\Omega_A$ around ${\bf \Omega}_A$ rewrites the
equations in the rotating frame where the $z-$axis is chosen along ${\bf
  \Omega}_A$. In this rotating frame, the acceleration (or the vector ${\bf
  u}$) is confined to the $xz$--plane and makes an angle $\alpha$ with the
$z$--axis.  Second, a rotation of angle $\alpha$ around the $y$--axis makes
the direction of acceleration along $z$. The corresponding transformations are
simple to carry out and the resulting equations are:
\begin{eqnarray}
\ddot{X}-2\Omega_A\, \cos\alpha\, \dot
Y&=&\Omega_A^2(\cos^2\alpha\,X+\sin\alpha\cos\alpha\, Z)\nonumber \\
&&-\frac{kX}{(X^2+Y^2+Z^2)^{3/2}},\\
\ddot{Y}+2\Omega_A\, (\cos\alpha\, \dot
X+\sin\alpha\, \dot Z)&=&\Omega_A^2 Y-\frac{kY}{(X^2+Y^2+Z^2)^{3/2}},\\
\ddot{Z}-2\Omega_A\, \sin\alpha\, \dot
Y&=&\Omega_A^2(\sin\alpha\cos\alpha\,X+\sin^2\alpha\,
Z)\nonumber \\&&-\frac{kZ}{(X^2+Y^2+Z^2)^{3/2}} + A.
\end{eqnarray}
With respect to the AKP, precession introduces the Coriolis force (second term
on the left hand side of each equation) and the centrifugal force with respect
to the rotation vector ${\bf \Omega}_A$ (first term on the right hand side of
each equation). The momenta of the precessing AKP are given as:
\begin{eqnarray}
p_X&=&\dot{X}-\Omega_A\, Y\, \cos\alpha,\\
p_Y&=&\dot{Y}+\Omega_A\, (X\,\cos\alpha+Z\,\sin\alpha),\\
p_Z&=&\dot{Z}-\Omega_A\, Y\,\sin\alpha,
\end{eqnarray}
and the corresponding Hamiltonian is:

\begin{eqnarray}
H&=&\frac{p^2}{2}+\Omega_A Y (p_X\cos\alpha +p_Z\sin\alpha)\nonumber \\&& -\Omega_A p_Y(X
\cos\alpha+Z\sin\alpha)-\frac{k}{|{\bf X}|}-AZ.
\end{eqnarray}
This in turn shows that the energy integral of the AKP is:
\begin{equation}
E=\frac{\dot X^2+\dot Y^2+\dot Z^2}{2}-\frac{\Omega_A^2}{2}\left[(X\cos\alpha+
Z\sin\alpha)^2+Y^2\right]-\frac{k}{|{\bf X}|}-AZ.
\end{equation}
As axisymmetry has now been lost, the vertical component of angular momentum
is not conserved. We have searched for analytic integrals of the precessing
AKP without much success. We note however that averaged precessing AKP is
integrable and that an exact analytical solution was derived in Paper I.
Further numerical investigations will shed some light on the integrability of
the precessing AKP.


\begin{thebibliography}{}




\bibitem{bj89}

  B.H. Bransden and C.J. Joachain, Introduction to Quantum Mechanics. Addison
  Wesley Longman Limited. (1989)

\bibitem{ep16}

  P.S. Epstein, Zur Theorie des Starkeffektes, Annalen der Physik {\bf 50},
  489 (1916)

\bibitem{ll60}

  L.D. Landau and E. M. Lifshitz, Course of Theoretical Physics. Pergamon
  Press Oxford. (1969)

\bibitem{n1}
  
  F. Namouni, On the origin of the eccentricities of extrasolar planets, AJ
  {\bf 130}, 280 (2005) (Paper I)

\bibitem{n2}
  
  F. Namouni, J.L. Zhou, The influence of mutual perturbations on the
  eccentricity excitation by jet acceleration in extrasolar planetary systems.
  CMDA {\bf 95}, 245 (2006)

\bibitem{n3}
  
  F. Namouni, On the flaring of jet-sustaining accretion disks. ApJ {\bf 659},
  1510  (2007) (Paper II)

\bibitem{r63}  
  P.J. Redmond, Generalization of the Runge-Lenz Vector in the presence of an
  electric field, Physical Review 133, B1352 (1964)

\bibitem{som}

  A. Sommerfeld, Atombau und Spektrallinen. Braunschweig: Vieweg \& Sohn.
  (1929)

\end{thebibliography}
\end{document}